\documentclass[12pt]{article}
\pdfoutput=1

\usepackage{amsmath,amssymb,amscd}
\usepackage{listings}
\usepackage{caption}
\usepackage{dsfont}
\usepackage{slashed}
\usepackage{color}

\usepackage[pdftex]{graphicx}
\usepackage{epstopdf}
\usepackage{subfigure}
\usepackage{epsfig}
\usepackage{listings}
\usepackage{caption}
\usepackage{cite}

\usepackage{multirow}

\newcommand{\bea}{\begin{align}}
\newcommand{\eea}{\end{align}}
\newcommand{\beq}{\begin{equation}}
\newcommand{\eeq}{\end{equation}}
\newcommand{\nbea}{\begin{align*}}
\newcommand{\neea}{\end{align*}}
\newcommand{\nbeq}{\begin{equation*}}
\newcommand{\neeq}{\end{equation*}}
\newcommand{\bear}{\begin{eqnarray}}  
\newcommand{\eear}{\end{eqnarray}}  

\numberwithin{equation}{section}

\begin{document}

\def\thefootnote{\fnsymbol{footnote}}

\begin{flushright}
{\tt KCL-PH-TH/2025-22, CERN-TH-2025-115}
\end{flushright}

\vspace{0.7cm}
\begin{center}
{\bf {\Large Personal Memories of 50 Years of Quarkonia}}
\end{center}

\medskip

\begin{center}{\large
{\bf John~Ellis}}
\end{center}

\begin{center}
{Theoretical Particle Physics and Cosmology Group, Department of
  Physics, King's~College~London, London WC2R 2LS, United Kingdom;\\
Theory Division, CERN, CH-1211 Geneva 23,
  Switzerland}
  \end{center}

\vspace{0.7cm}

\centerline{\bf ABSTRACT}
~~\\
\noindent  
The world of particle physics was revolutionised in November 1974 by the discovery of the J/$\psi$ particle, the first particle to be identified as a quarkonium state composed of charm quarks and antiquarks. The charmonium interpretation of the J/$\psi$ was cemented by the subsequent observations of a spectrum of related $\bar c c$ states, and finally by the discovery of charmed particles in 1976. The discovery of charmonium was followed in 1977 by the discovery of bottomonium mesons and particles containing bottom quarks. Toponium bound states of top quark and antiquarks were predicted to exist in principle but, following the discovery of the top quark in 1995, most physicists thought that its observation would have to wait for a next-generation $e^+ e^-$ collider. However, in the second half of 2024 the CMS Collaboration reported an excess of events near the threshold for $\bar t t$ production at the LHC that is most plausibly interpreted as the lowest-lying toponium state. These are the personal recollections of an eyewitness who followed closely these 50 years of quarkonium discoveries.\\
~~\\
Contribution to the Proceedings of the Symposium held at IHEP, Beijing on October 20th, 2024 to celebrate the 50th Anniversary of the Discovery of the J/$\psi$ Particle.\\
~~\\
{\it This paper is dedicated to the memory of Mary~K.~Gaillard, with whom I was privileged to share many happy collaborations thinking about heavy quarks and their quarkonia.}

\vspace{0.7cm}
\begin{flushleft}
June 2025
\end{flushleft}

\newpage

\section{Prehistory}

In hindsight, the quarkonium story can be thought to have begun in 1963 with the discovery of the $\phi$ particle~\cite{phi}, a particle lying just above the $K \bar K$ threshold which nevertheless preferred decaying into kaons rather than the relatively light, kinematically-preferred pions. This preference was soon interpreted by Susumu Okubo~\cite{Okubo} as a consequence of approximate SU(3) flavour symmetry.  Then, at the end of 1963 and the beginning of 1964, Andr{\'e} Petermann~\cite{Petermann}, Murray Gell-Mann~\cite{Gell-Mann} and George Zweig~\cite{Zweig} proposed independently that hadrons might be composed of elementary constituents, that Gell-Mann called quarks (the name that stuck) and Zweig called ``aces”. In his work, Zweig suggested that the $\phi$ was a bound state of a strange quark-antiquark pair, and proposed that this could explain the preferred coupling of the $\phi$ to the strange kaons via the Zweig (also called the Okubo-Zweig-Iizuka~\cite{Iizuka} (OZI)) rule, according to which hadronic decays are suppressed if the initial quark–antiquark pair must be annihilated. After 1974 the portmanteau word “strangeonium” was retrospectively applied to the $\phi$ and similar heavier $s \bar s$ bound states, but the name has never really caught on.

In the year or so prior to the J/$\psi$ discovery, there was much speculation about data from the CEA~\cite{CEA} and SPEAR~\cite{SPEAR} $e^+ e^-$ colliders that indicated a rise in the ratio, $R$, of the cross-sections for hadron production relative to that for $\mu^+ \mu^-$ pairs. Was this a failure of the parton model that had only recently found acceptance as a model for scaling in deep-inelastic scattering? Or did partons have internal structure? Or were there ``new” partons that had not been seen previously, such as charm or integrally-charged coloured quarks? 

I was asked on several occasions to review the dozens of theoretical suggestions on the market, e.g., at the ICHEP conference in the Summer of 1974~\cite{Ellis1974} (see also~\cite{CE,RAL}). In preparation, I toted around Europe a heavy-duty Migros supermarket shopping bag filled with dozens of theoretical papers. Playing the part of an objective reviewer, I did not come out strongly in favour of any specific interpretation. However, during talks that Autumn in Copenhagen and Dublin I finally spoke out in favour of charm as the best-motivated explanation of the increase in $R$.

\section{Charmonium}

Then, on November 11th 1974 the news broke that two experimental groups, one working at BNL under the leadership of Sam Ting~\cite{J} and the other at SLAC~\cite{psi} led by Burt Richter, had discovered in parallel the narrow vector boson that bears the double-barrelled name J/$\psi$. The worldwide particle physics community went into convulsions, and the CERN Theory Division was no exception. We held informal midnight discussion sessions around an open-mic phone with Fred Gilman in the SLAC theory group, who generously shared with us the latest J/$\psi$ news. Away from the phone, like groups around the world, we debated the merits and demerits of many different theoretical ideas. Instead of writing a plethora of individual papers about these ideas, we decided to bundle our preliminary thoughts into a collective preprint~\cite{CoCo}. Rather than claim credit for our elementary thoughts, the preprints' authors were anonymous, the place of the authors’ names being taken by a mysterious ``CERN Theory Boson Workshop”.~\footnote{Eagle eyes will spot that the equations were handwritten by Mary K. Gaillard. Informally, we called ourselves ``Co-Co”, for ``Communication Collective”.}

My immediate personal instinct was to advocate the charmonium interpretation of the J/$\psi$, on the basis of the motivation from the Sheldon Glashow, John Iliopoulos and Luciano
Maiani (GIM)~\cite{GIM} mechanism for the suppression of flavour-changing neutral currents, which led to the estimate by Mary K Gaillard and Ben Lee that the charm quark should weight $\sim 2$~GeV~\cite{GL} (see also~\cite{GLR}) and invoking the Zweig (or OZI) rule to suppress charmonium hadronic decays. This interpretation of the J/$\psi$ was strongly supported by the work of Tom Appelquist and David Politzer~\cite{AP} (see also~\cite{DG}), who showed that charmonium decay could be calculated semi-quantitatively within QCD, thanks to asymptotic freedom, and was the first interpretation to be described in our paper. Nevertheless, one of the authors of the GIM paper co-authored a paper suggesting that the J/$\psi$ might be an intermediate electroweak vector boson~\cite{Z}. 

A few days after the J/$\psi$ discovery came the news of the discovery at SLAC of the (almost equally narrow) $\psi^{\prime}$~\cite{Abrams}, which I was told as I was walking along the theory corridor to my office one morning. My informant was a senior CERN theorist who was convinced that this discovery would kill the charmonium interpretation of the J/$\psi$. However, before I reached my office I realised that the $\psi^\prime$ could be a radial excitation of charmonium, and that an extension of the Zweig rule would also suppress $\psi^{\prime} \rightarrow \, $J/$\psi +$ light meson decays, so the $\psi^{\prime}$ could also be narrow.

As was pointed out soon after by Appelquist, Alvaro De Rujula, Politzer and Glashow~\cite{ADPG} that the charmonium interpretation of the J/$\psi$ and $\psi^{\prime}$ states predicted that there should be intermediate P-wave states (in which the charm quark and antiquark would carry orbital angular momentum) that could be detected in radiative decays of the $\psi^{\prime}$. In the first half of 1975 there was keen competition between teams at SLAC and DESY to discover these states. That summer I was visiting SLAC, where one day I discovered under the cover of a copying machine, before their discovery was announced, a sheet of paper with plots showing clear evidence for P-wave states. I made a copy, went straight to Burt Richter's office and handed him the sheet of paper. I also asked whether he wanted the copy I had made. He graciously allowed me to keep it, as long as I kept quiet about it, which I did until the discovery was officially announced a few weeks later~\cite{DASPP,SLACP}.

Discussion about the interpretation of the new particles, in particular between advocates of charm and coloured quarks, rumbled on for a couple of years until the discovery of charmed particles in 1976~\cite{Goldhaber:1976xn}. During this period we conducted some debates in the main CERN auditorium moderated by John Bell. I remember one such debate in particular, during which a distinguished senior British theorist spoke for integer-charged coloured quarks~\cite{FM} and I spoke for charm. I was a bit surprised when he described me as representing the ``Establishment", as I was under 30 at the time.

\section{Bottomonium}

The discovery of the $\tau$ lepton in 1975~\cite{tau} suggested strongly that there should exist an additional pair of heavier quarks, called top and bottom, which would have the benefit of accommodating CP violation in an economical way~\cite{KM}. However, at the time there were no convincing arguments what the masses of these quarks might be.

Over the following year my attention wandered to grand unified theories that unify the electroweak and strong interactions within a larger symmetry group appearing at high energies. My first paper on the subject was with Michael Chanowitz and Mary K. Gaillard~\cite{CEG}, which we completed in May 1977. We realised while writing this paper that simple grand unified theories would relate the mass of the $\tau$ heavy lepton to that of the bottom quark. Our prediction was $m_b/m_\tau = 2~{\rm to}~5$, which we included in the main text, but not in the Abstract. Shortly afterwards, while our paper was in proof, the discovery of the $\Upsilon$ particle~\cite{LL} by a group at Fermilab led by Leon Lederman became known, implying that $m_b \sim 4.5$~GeV. I added our successful mass prediction by hand in the margin of the corrected proof. Unfortunately, the journal misunderstood my handwriting and printed our prediction as $m_b/m_\tau = 2605$, a spectacularly inaccurate postdiction!

Meanwhile, buoyed by the success of our prediction for $m_b$, Mary K. Gaillard, Dimitri Nanopoulos, Serge Rudaz and I set to work on a paper about the phenomenology of the top and bottom quarks~\cite{EGNR}. One of our predictions was that the first two excited states of the $\Upsilon$, the $\Upsilon^{\prime}$ and $\Upsilon^{\prime \prime}$ should be detectable in the Lederman experiment because the Zweig rule would suppress their cascade decays to lighter bottomonia via light meson emission. Indeed, the Lederman experiment found that the $\Upsilon$ bump was broader than the experimental resolution, and the bump was eventually resolved into three resonance peaks~\cite{Innes}.

It was in the same paper that we introduced the terminology of ``penguin diagrams”, loop-level processes that can generate rare decay modes, CP violation, and effects sensitive to new physics beyond the Standard Model. Similar diagrams had been discussed by the Moscow ITEP theoretical school~\cite{Shifman} in connection with $K$ decays, and we realised that they would be important in $B$ hadron decays. As has been described elsewhere, I took an evening off to go to a bar in the Old Town of Geneva, where I got involved in a game of darts with Melissa Franklin. She bet me that if I lost the game I had to include the word ``penguin” in my next paper. Melissa actually abandoned the darts game before the end, and was replaced by Serge Rudaz, who beat me. I still felt obligated to carry out the conditions of the bet, but for some time it was not clear to me how to get the word ``penguin" into the paper about $b$ and $t$ quarks that we were writing at the time. Then, another evening after working at CERN I stopped to visit some friends on my way back to my apartment, and inhaled some (at that time) illegal substance. Later, when I got back to my apartment and continued working on our paper, I had a sudden inspiration that the famous Russian diagrams look like penguins. So we put the word into our paper, and it has now appeared in almost 10000 papers.

\section{Toponium?}

What of toponium, the last remaining frontier in the world of quarkonia? In the early 1980s there were no experimental indications how heavy the top quark might be, and there were hopes that it might be within range of existing or planned $e^+ e^-$ colliders such as PETRA, TRISTAN and LEP. When the LEP experimental programme was being devised, candidate experimental designs I was involved in setting a number of ``examination questions" for them that included asking how well they could measure the properties of toponium. In parallel, the first theoretical papers on the formalism for toponium production in $e^+ e^-$ and hadron-hadron collisions appeared~\cite{FK,FKS}. 

But the top quark did not appear until the mid-1990s at the Fermilab proton-antiproton collider, with a mass around 175 GeV~\cite{CDF,D0}, implying that toponium measurements would require an $e^+ e^-$ collider with an energy much greater than LEP, around 350~GeV. Many theoretical studies were made of the cross section around the mass of vector (ortho)toponium and in the neighbourhood of the $e^+ e^- \to t \bar t$ threshold, and how precisely the top quark mass, electroweak and Higgs couplings could be measured, see for example~\cite{Vos}.

Meanwhile, a number of theorists have been calculating the possible toponium signal at the LHC~\cite{Kiyo,Sumino,Ju,Fuks1,Aguilar,Fuks,Garzelli}, and in recent years the LHC experiments ATLAS and CMS have been measuring $\bar t t$ production with high statistics. CMS and ATLAS embarked on programmes to search for quantum-mechanical entanglement in the final-state decay products of the t quarks and antiquarks, as should occur if the $\bar t t$ state were to be produced in a specific spin-parity state~\cite{ATLASQM,CMSPS}. They both found decay correlations characteristic of $\bar t t$ production in a pseudoscalar state, the first time quantum entanglement had been observed at such high energies.

The CMS Collaboration used these studies to improve the sensitivities of dedicated searches they were making for possible heavy Higgs bosons decaying into $\bar t t$ final states, as would be expected in many extensions of the Standard Model~\cite{CMSPAS}. Intriguingly, hints of a possible excess of events around the $\bar t t$ threshold with the type of entanglement expected in a pseudoscalar $\bar t t$ state began to emerge in the CMS data, but initially not with high significance. 

Rumours of this excess first reached me at an Asia-CERN physics school in Thailand in June 2024. Motivated by the need to prepare a lecture on top physics for a subsequent physics school, I started wondering whether the CMS excess could be due to a heavy pseudoscalar Higgs boson or to the lowest-lying $^1S_0$  (para)toponium bound state, nestling in the colour-singlet Coulomb-like $\bar t t$ potential, and how one might distinguish between these hypotheses. As longstanding protagonists of heavy Higgs bosons, a few years previously Abdelhak Djouadi, Andrei Popov, J{\' e}r{\' e}mie Quevillon and I had studied in detail the possible signatures of heavy Higgs bosons in $\bar t t$ final states at the LHC~\cite{DEPQ}. We had shown that they could have significant interference effects with the QCD background that would generate dips in the cross-section as well as bumps, that could modify substantially the production cross-section. 
\textcolor{red}{Motivated by the CMS excess, Abdelhak, J{\' e}r{\' e}mie and I studied some more how the toponium and Higgs hypotheses could be distinguished~\cite{DEQ}.} 

Subsequently the significance of the CMS signal increased to over $5\sigma$ in a tailored search for new pseudoscalar states decaying into $\bar t t$ pairs with specific pseudoscalar spin correlations~\cite{CMSXS}, \textcolor{red}{and this CMS discovery has subsequently been confirmed by the ATLAS Collaboration, with a significance over $7\sigma$~\cite{ATLASXS}.} Unfortunately, the experimental resolution in the $\bar t t$ invariant mass is not precise enough~\cite{CMSAH} to see any dip as would appear in pseudoscalar Higgs production~\cite{DEQ}, so it is not (yet) possible to distinguish between the toponium and Higgs hypotheses on purely experimental grounds.

Despite being a fan of extra Higgs bosons, I have to concede that toponium is the more plausible interpretation of the \textcolor{red}{$\bar t t$} threshold excess. The mass is consistent with that expected for toponium, the signal strength is consistent with theoretical calculations in QCD, and the $t \bar t$ spin correlations are just what one expect for the lowest-lying pseudoscalar (para)toponium $^1S_0$ state that would be produced in gluon-gluon collisions. 

\section{The future?}

Caution is still in order. \textcolor{red}{The pseudoscalar Higgs hypothesis cannot (yet) be excluded.} Nevertheless, it would be a wonderful Golden Anniversary present for quarkonium if, 50 years after the discovery of the J/$\psi$, the appearance of its last, most massive sibling were to be confirmed.

Toponium will be a very interesting target for future $e^+ e^-$ colliders, which will be able determine its properties with much greater accuracy than the LHC could ever achieve, making possible precise measurements of the mass of the top quark and its electroweak couplings. The quarkonium saga is far from over.


\section*{Acknowledgements}

This work was supported by the STFC (UK) via the Research Grant ST/T000759/1.


\end{document}